# Emittance Minimization for Aberration Correction I: Aberration correction of an electron microscope without knowing the aberration coefficients

Desheng Ma*

*School of Applied and Engineering Physics,*
*Cornell University, Ithaca, NY 14853, USA*

Steven E. Zeltmann

*Platform for the Accelerated Realization, Analysis, and Discovery of Interface Materials*
*and*
*School of Applied and Engineering Physics,*
*Cornell University, Ithaca, NY 14853, USA*

Chenyu Zhang and Zhaslan Baraissov

*School of Applied and Engineering Physics,*
*Cornell University, Ithaca, NY 14853, USA*

Yu-Tsun Shao

*Mork Family Department of Chemical Engineering and Materials Science,*
*University of Southern California, Los Angeles, CA 90089, USA*

Cameron Duncan and Jared Maxson

*Department of Physics, Cornell University, Ithaca, NY 14853, USA*

Auralee Edelen

*SLAC National Accelerator Laboratory, Menlo Park, CA 94025, USA*

David A. Muller[†]

*School of Applied and Engineering Physics,*
*Cornell University, Ithaca, NY 14853, USA and*
*Kavli Institute at Cornell for Nanoscale Science, Ithaca, NY 14853, USA*

(Dated: November 24, 2024)

* Funded by the Center for Bright Beams, an NSF STC (NSF PHY-1549132).

† dm852@cornell.edu

‡ dm24@cornell.edu

# Abstract

Precise alignment of the electron beam is critical for successful application of scanning transmission electron microscopes (STEM) to understanding materials at atomic level. Despite the success of aberration correctors, aberration correction is still a complex process. Here we approach aberration correction from the perspective of accelerator physics and show it is equivalent to minimizing the emittance growth of the beam, the span of the phase space distribution of the probe. We train a deep learning model to predict emittance growth from experimentally accessible Ronchigrams. Both simulation and experimental results show the model can capture the emittance variation with aberration coefficients accurately. We further demonstrate the model can act as a fast-executing function for the global optimization of the lens parameters. Our approach enables new ways to quickly quantify and automate aberration correction that takes advantage of the rapid measurements possible with high-speed electron cameras. In part II of the paper, we demonstrate how the emittance metric enables rapid online tuning of the aberration corrector using Bayesian optimization.

## I. INTRODUCTION

Electron lenses are by their very nature imperfect. The magnetic solenoid lenses ubiquitously used in electron microscopy not only suffer from distortions due to manufacturing imperfections, but also possess a spherical aberration which Scherzer proved cannot be made to vanish by careful design [1]. While stigmators were quickly developed to correct many of these optical distortions, the spherical aberration long limited the resolution of electron microscopes. Correction of electron lens aberrations was long understood to be possible using non-round lens elements in increasingly complex arrangements [2], but it was not until the turn of the 21st century that these efforts succeeded in improving the resolution of the imaging system [3–5]. The development of spherical aberration ($C_s$) correctors (which actually correct a wider range of aberrations) [6–10], and critically of online methods for characterizing the aberrations of the system, brought sub-ångström resolution in both conventional and scanning transmission electron microscopy (TEM).

Historical efforts at aberration correction were often hampered by the lack of systematic procedures for aligning the dozens of optical elements they added to the system. While modern correctors with computer control have vastly improved on this situation, tuning of the aberration corrector remains a complex and time-consuming process. The reasons are threefold: (i) The experimental aberration function is unknown. Existing methods can only provide estimates of the aberration coefficients and leave higher order residues [9,11–19]. (ii) Even if the aberration function were exactly known, different orders of aberrations are coupled and difficult to reduce at the same time [20,21]. Individual electromagnetic elements in the corrector can each influence many aberrations, so mapping from measured aberration coefficients to an optimal set of currents in the coils is difficult. (iii) The observed lifetime of a well corrected state is short (less than an hour) and suffers from intrinsic instability as a fundamental limit, necessitating constant tuning [21–24]. Rose summarized the situation and pointed out that the aberrations must be "eliminated within a period of time which must be shorter than the duration of the overall stability of the entire system" [3].

There are currently two major classes of methods to measure aberrations, which rely on analyzing either incoherent images in real space or Ronchigrams of coherent diffraction patterns. In the real space approach, a series of images of a calibration specimen are acquired at different tilt angles. The shifts of these images and the deconvolved probe shape are used to determine, respectively, the first and second derivatives of the aberration function at each illumination tilt [11–16]. This approach is commonly seen on microscopes equipped with CEOS correctors, and the number of images required (and thus the acquisition time) will depend on the number of aberration coefficients to be measured – measuring sets of higher order coefficients requires many more images. For aberration measurements from diffraction patterns, Ronchigrams are acquired and segmented, with an intentionally manipulated parameter once at a time, e.g., a known defocus change or beam shift, then cross-correlated to fit to the aberration function [9,18,19,25]. This approach was originally developed by Nion and later adopted by JEOL. More recently, rapid aberration measurements were demonstrated using artificial neural networks to diagnose Ronchigrams [26,27]. Ultimately, from the perspective of tuning the optics, none of these approaches directly yield the optimal change in currents run in the electromagnetic lenses. While the aberration function provides an interpretable representation of the lens-induced distortions, the estimated aberration coefficients still need to be converted back to lens currents by a calibrated mapping. This mapping in practice can drift from day to day and may only be valid over a small range of misalignments, necessitating many iterations of diagnosis and correction.

In this work, we consider a different approach to electron optical diagnosis and tuning that bypasses the need to measure individual aberrations in the microscope alignment task. An ideal metric for use in an optimization routine should be single valued, convex, and with a lower bound at the optimally aligned state. Some proposed metrics include the Strehl ratio measured from the Ronchigram [27] and the normalized image variance on a featureful specimen [28–30]. However, the experimental implementation of these metrics remains largely heuristic. Instead, we find such a metric by drawing connections with the field of accelerator physics: an electron microscope is essentially a room-sized particle accelerator and in aberration correction we ultimately share the same goal as preservation of beam brightness by minimizing beam emittance, a physical quantity defined as the volume the beam occupies in the phase space [31–34]. We derive the emittance growth of an electron beam introduced by aberrations by transforming the 2D wave function defined by the aperture function and the aberration function into a phase space via the Wigner-Weyl transform. Furthermore, we show the derived beam emittance growth is convex with respect to aberration coefficients. The emittance is therefore a superior beam quality metric than the aberration function in the sense that it is single valued but contains all phase space information to fully characterize the beam with a guaranteed stable optimum. Now the question becomes: how do we measure beam emittance growth in an electron microscope?

While 4D-STEM, which directly records scattering information in a position-momentum space, appears to be attractive approach to measuring emittance, processing of this phase space information is relatively slow and thus inefficient for online tuning [17,35]. Instead, we create a supervised deep learning model to build an accurate and fast-executing mapping from individual Ronchigrams to emittance growth. Deep neural networks have shown good performance in image analysis and pattern recognition, including recent applications in electron microscopy, e.g., diffraction analysis [36], convergence angle selection [27] and 4D-STEM data processing [37]. We develop a convolutional neural network based on an award-winning architecture to cater to our Ronchigram-emittance growth regression problem. This machine learning approach is validated on both a simulation study and real experiments on aberration-corrected STEMs. This approach not only builds a reliable nonlinear mapping from Ronchigrams to beam emittance growth, but also allows for usage as the surrogate function for fast online global optimization of the lens parameters to achieve aberration correction. In Part II of the paper, we will demonstrate how this neural network can be utilized to can achieve rapid online tuning of the aberration corrector using Bayesian optimization.

## II. METHODOLOGY

We introduce our methodology in two sections: (i) Definition of *beam emittance*; (ii) Architecture of a customized Convolutional Neural Network (CNN) to predict beam emittance growth from Ronchigrams. We emphasize that for the purpose of aberration correction we need not know the full emittance of the beam, which also includes a contribution from the beam source, but only the extra emittance growth introduced by aberrations. In the rest of the paper, the two terms *emittance* and *emittance growth* are sometimes used interchangeably.

### A. Beam emittance growth from aberration coefficients

Electron-optical alignment problems are not unique to electron microscopes. Particle accelerators such as synchrotron rings use similar multipole optical elements, including quadrupoles and sextupoles. In accelerator physics, emittance is a property of a charged particle beam and is defined as the volume occupied by the beam in the phase space, which according to Liouville's Theorem, is a conserved quantity under conservative forces. In practice, we can apply the statistical (root-mean-squared) definition of beam emittance given by,

$$\varepsilon = \sqrt{\langle \vec{\xi}^2 \rangle \langle \vec{\alpha}^2 \rangle - \langle \vec{\xi} \cdot \vec{\alpha} \rangle^2} = \sqrt{\det(\text{cov}(\vec{\xi}, \vec{\alpha}))} \quad (1)$$

where $\vec{\xi} = (x, y)$ is the position and $\vec{p} \approx (x', y') \approx \vec{\alpha}$ is the momentum (angle) in the 4D subspace orthogonal to the optical axis under paraxial (Gaussian-optics) approximation, and $\text{cov}(\vec{\xi}, \vec{\alpha}) = \mathbf{C} = \begin{bmatrix} \langle \vec{\xi}^2 \rangle & \langle \vec{\xi} \cdot \vec{\alpha} \rangle \\ \langle \vec{\xi} \cdot \vec{\alpha} \rangle & \langle \vec{\alpha}^2 \rangle \end{bmatrix}$ is the covariance matrix between the two. For an electron microscope, emittance determines the source brightness $B = I/\varepsilon^2$ where $I$ is the incident current. This definition is classical and is commonly used in particle-tracing simulations [32–34]. A quantum equivalent defined with quantum operators is discussed in Appendix D.

The standard expression to describe beam quality due to imperfect lenses within the field of electron microscopy is the aberration function. Here we use the Krivanek notation [1] where the phase shift caused by imperfect lenses can be expanded in terms of the radial ($\alpha$) and azimuthal ($\phi$) angles as

$$\begin{aligned}
\chi(\alpha, \phi) &= \frac{2\pi}{\lambda} \sum_{n,m} \frac{C_{n,m} \alpha^{n+1} \cos\left(m(\phi - \phi_{n,m})\right)}{n+1} \\
&= \frac{2\pi}{\lambda} \sum_{n,m} \frac{\alpha^{n+1}}{n+1} [C_{nma} \cos(m\phi) + C_{nmb} \sin(m\phi)] \\
&= \frac{2\pi}{\lambda} \sum_{n,m} \frac{\alpha^{n+1}}{n+1} \Re[\tilde{C}^*_{n,m} e^{im\phi}]
\end{aligned} \quad (2)$$

where $n$ is the order of aberration and $m$ the order of rotational symmetry. A simple model of the resolution $\delta$ beyond the Rayleigh limit for a rotationally-symmetric system can be defined by a polynomial of $\alpha$ with the aberration coefficients, in the simplest case as follows [24],

$$\delta^2 = (0.61\lambda/\alpha)^2 + (C_1\alpha)^2 + (C_3\alpha^3)^2 + (C_5\alpha^5)^2 + \cdots \quad (3)$$

Equation 3 implies there is an intrinsic instability of the corrected state of the probe at the optimal resolution $\delta_0$ given by $d\delta/d\alpha = 0$, as its derivatives with respect to aberration coefficients are diverging, e.g., $\partial \delta_0/\partial C_3 \sim 1/C_3^{3/4}$ and $\partial \delta_0/\partial C_5 \sim 1/C_5^{5/6}$. This cusp-like response indicates any small fluctuation of the aberration coefficients will perturb the corrected state, causing significant difficulty in online tuning the beam or maintaining its tuned state [24].

In a STEM, the 2D wave function of the electron probe is defined in the angular basis with basis vector, $\vec{a}$, by two real functions: the aperture function $\psi_0(\vec{\alpha}) = A(\vec{\alpha})$ which equals 1 inside the aperture cutoff of $\alpha_0$ and 0 outside, and the aberration function $\chi(\vec{\alpha})$,

$$\psi_{abr}(\vec{\alpha}) = \psi_0(\vec{\alpha}) e^{i\chi(\vec{\alpha})} \quad (4)$$

The Wigner-Weyl transform fills the gap between the expectation values in the classical representation of Equation 1 and the quantum representation of Equation 4, which maps linear operators in Hilbert space to the phase space [38], providing the quasi-probability of finding a particle at a given position $\vec{\xi}$, traveling with a given angle $\vec{\alpha}$.

$$\begin{array}{ccc}
\text{Hibert space operators} & \leftrightarrow & \text{phase space formulation} \\
\hat{O}(\hat{x}, \hat{\mathbf{p}}) = \dfrac{\hat{x}\hat{\mathbf{p}} + \hat{\mathbf{p}}\hat{x}}{2} & \leftrightarrow & O(\vec{\xi}, \vec{\alpha}) = \vec{\xi} \cdot \vec{\alpha} \\
\hat{\rho} = \psi_{abr} & \leftrightarrow & W(\vec{\xi}, \vec{\alpha})
\end{array} \quad (5)$$

The Wigner function (the Wigner quasi-probability distribution) is the Wigner transform of the density operator, which in our case of a pure state has the form,

$$W(\vec{\xi},\vec{\alpha}) = \frac{1}{4\pi^2}\int d\alpha'^2 \tilde{\psi}^*\left(\vec{\alpha}-\frac{1}{2}\vec{\alpha}'\right)\tilde{\psi}\left(\vec{\alpha}+\frac{1}{2}\vec{\alpha}'\right)e^{i\vec{\xi}\cdot\vec{\alpha}'} \tag{6}$$

The position and momentum space probabilities are given by the marginals:

$$\int d\xi^2 W(\vec{\xi},\vec{\alpha}) = |\tilde{\psi}(\vec{\alpha})|^2 \tag{7}$$

which is the STEM diffraction pattern, and

$$\int d\alpha^2 W(\vec{\xi},\vec{\alpha}) = |\tilde{\psi}(\vec{\xi})|^2 \tag{8}$$

which is the point spread function or probe shape in real space.

Although with the Wigner function we can calculate the second moments for Equation 1 directly, we realize a cleaner expression is obtainable with careful inspection upon substituting Equation 4 into Equation 6.

$$\begin{aligned} W(\vec{\xi},\vec{\alpha}) &= \frac{1}{4\pi^2}\int d\alpha^2 \psi^*\left(\vec{\alpha}-\frac{1}{2}\vec{\alpha}'\right)\psi\left(\vec{\alpha}+\frac{1}{2}\vec{\alpha}'\right)e^{i\vec{\xi}\cdot\vec{\alpha}'} \\ &= \frac{1}{4\pi^2}\int d\alpha'^2 \psi_0^*\left(\vec{\alpha}-\frac{1}{2}\vec{\alpha}'\right)\psi_0\left(\vec{\alpha}+\frac{1}{2}\vec{\alpha}'\right)e^{i\chi(\vec{\alpha}-\frac{1}{2}\vec{\alpha}')-i\chi(\vec{\alpha}+\frac{1}{2}\vec{\alpha}')}e^{i\vec{\xi}\cdot\vec{\alpha}'} \\ &\approx \frac{1}{4\pi^2}\int d\alpha'^2 \psi_0^*\left(\vec{\alpha}-\frac{1}{2}\vec{\alpha}'\right)\psi_0\left(\vec{\alpha}+\frac{1}{2}\vec{\alpha}'\right)e^{i(\vec{\xi}+\nabla\chi(\vec{\alpha}))\cdot\vec{\alpha}'} \end{aligned} \tag{9}$$

Comparing to the Wigner distribution without aberrations, Equation 9 suggests that under first order approximation the aberration only adds an angle dependent real space shift as shown in Equation 10 to the probe.

$$\Delta\vec{\xi} \approx \nabla\chi(\vec{\alpha}) \tag{10}$$

where $\chi$ is the aberration function with angles as variables. This together with Equation 1 leads us to an expression for the terms of the emittance which are introduced by aberrations:

$$\begin{aligned} \varepsilon_\chi^2 &= \langle\nabla\chi^2\rangle\langle\vec{\alpha}^2\rangle - \langle\nabla\chi\cdot\vec{\alpha}\rangle^2 \\ &= \left(\int d\alpha^2 A^2(\vec{\alpha})|\vec{\nabla}\chi(\vec{\alpha})|^2\right)\left(\int d\alpha^2 A^2(\vec{\alpha})|\vec{\alpha}|^2\right) - \int d\alpha^2 A^2(\vec{\alpha})\vec{\alpha}\cdot\nabla\chi(\vec{\alpha}) \end{aligned} \tag{11}$$

Consider a perfect aperture $A^2(\alpha,\phi) = 1/\pi\alpha_0^2$ for $\alpha < \alpha_0$, substituting into Equation (11) yields,

$$\varepsilon_\chi^2 = \frac{4\pi^2}{\lambda^2}\sum_{nn'm}\Re[\tilde{C}_{n,m}^*\tilde{C}_{n',m}]\alpha_0^{n+n'+2}\left(\frac{1}{n+n'+2} + \frac{m^2}{(n+1)(n'+1)(n+n'+2)} - \frac{4\delta_{m0}}{(n+3)(n'+3)}\right) \tag{12}$$

Now we can examine the properties of the derived emittance growth induced by aberrations.

**Lemma 1.** The emittance is independent of defocus:

$$\frac{d\varepsilon_\chi^2}{d\tilde{C}_{10}} = 0 \tag{13}$$

This result directly follows from the translational invariance of the optical system as a perfect lens introduces a size-preserving shear in phase space.

**Lemma 2.** The emittance growth is convex in aberration coefficients.

$$\frac{d^2\varepsilon_\chi^2}{d\tilde{C}_{nm}d\tilde{C}_{n'm}} = \alpha_0^{n+n'+2}\left(\frac{1}{n+n'+2} - \frac{4\delta_{m0}}{(n+3)(n'+3)} + \frac{m^2}{(n+1)(n'+1)(n+n'+2)}\right) \tag{14}$$

where the Hessian is constant and positive semi-definite. This implies the derived $\varepsilon_\chi^2$ not only guarantees a stable optimum, but also can be directly optimized using Newton's method.

We can rewrite Equation 11 as follows,

$$\varepsilon_\chi = \sqrt{\langle\nabla\chi^2\rangle\langle\vec{\alpha}^2\rangle - \langle\nabla\chi \cdot \vec{\alpha}\rangle^2} = \sqrt{\det(\text{cov}(\nabla\chi, \vec{\alpha}))} \tag{15}$$

This expression allows a much more efficient way to calculate emittance growth, by using the numerical gradient of the aberration function $\chi$ over a Cartesian grid in $\vec{\alpha} = (x', y')$, then calculating the determinant of the covariance matrices.

In summary, beam emittance growth serves as a robust metric for aberration correction in electron microscopy because (i) It is a unified single-value metric, bypassing the need to correct individual aberration coefficients; (ii) It is bounded above 0 where zero corresponds to the aberration-free case. (iii) It does not change with defocus, allowing for greater flexibility in the tuning process (although defocus can be added to the error metric when needed -see below). (iv) It is convex in aberration coefficients and thus guarantees a stable optimum for rapid and robust online optimization. Figure 1 compares the simulated electron beam with higher and lower emittance growth values.

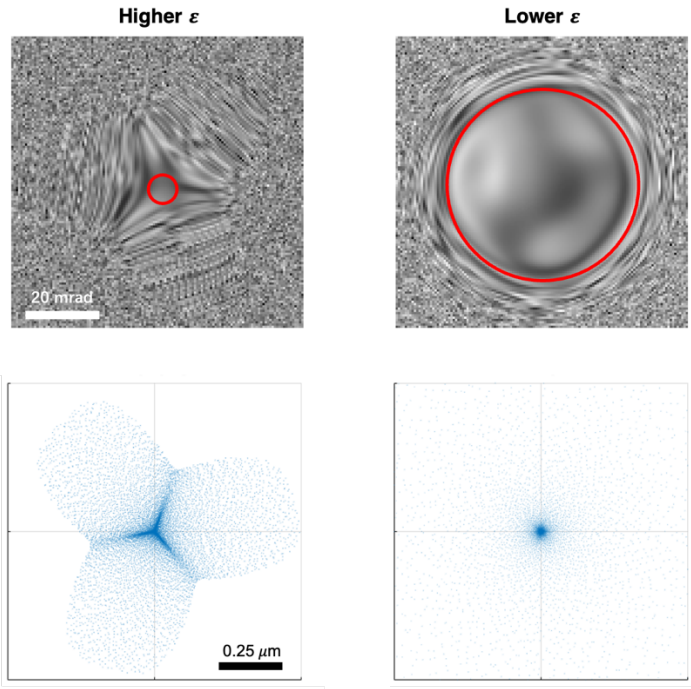

FIG. 1: High emittance (left) v.s. low emittance (right) beam. Top: Electron Ronchigrams. Bottom: Transverse distribution of electrons. The Ronchigram of lower emittance shows a much larger flat area in the center, corresponding to a more confined beam and larger usable angular range with low optical distortions.

### B. Customized convolutional neural network to predict emittance growth

Electron Ronchigrams contain all the information of phase shifts introduced by the lenses. Therefore, we can extract the beam emittance growth from the Ronchigram taken under each set of different lens parameters. This is a straight-forward image recognition task that has been extensively studied in the computer vision and deep learning communities.

We develop a CNN based on the ConvNet architecture VGG-16 [39] commonly used in image recognition, containing 16 weighted layers (13 convolutional layers and 3 fully connected (FC) layers) and 5 maxpooling layers. Unlike the original VGG-16 designed for image classification problems, we replace the last fully-connected layer of the model with a new one-dimensional linear layer instead of 1,000 in order to perform the regression task of predicting the continuous response of emittance. No activation is used for the final layers. Two additional dropout layers are inserted between the FC layers to prevent overfitting. The objective function is set to be the root mean squared error (RMSE) of the predicted emittance growth versus the ground truth calculated emittance growth from aberration coefficients. The Adam optimizer is used for optimization. The final CNN contains 16,812,353 parameters and is illustrated in Figure 2. The CNN is implemented in PyTorch [40].

Our CNN is trained on 25,000 simulated Ronchigrams under various types of aberrations labeled by their corresponding calculated probe beam emittance. The necessity of training on simulated data is due to three main reasons: (i) It is difficult to collect trustworthy labeled data from a real microscope, as the aberration coefficients are only estimated. (ii) The probe on a real electron microscope is intrinsically unstable (iii) The amount of training data needed for the neural network to converge is huge. Instead of training this deep neural network from scratch, which would require a significant amount of data, power and time, we pursue transfer learning to simplify and speed up the process by borrowing weights from a pretrained VGG-16 and finetune its performance with simulated Ronchigrams. We adopt a two-stage finetune training procedure, first only training the FC layers with all pretrained convolutional layers frozen by locking their gradient updates, then finetuning all layers by unlocking the gradient updates until convergence.

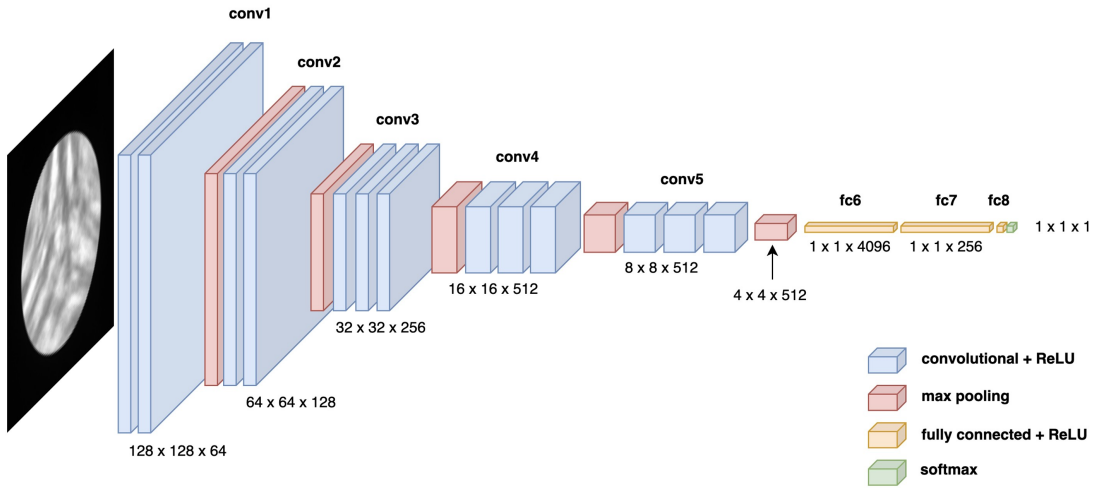

FIG. 2: VGG-16 architecture to map Ronchigrams to beam emittance. The neural network contains 16 convolutional layers where the first layer takes the Ronchigram image as input and propagates to the last layer which maps to a single value for beam emittance growth. Detailed hyperparameter settings are listed in Table III.

The training data set consists of Ronchigrams under various types of aberrations, generated by an open-source simulator [41]. The Ronchigram is the diffraction pattern of a convergent beam focused on an amorphous specimen and can be simulated from the probe wave function $\psi_{abr}(\vec{\alpha}) = \psi_0(\vec{\alpha})e^{i\chi(\vec{\alpha})}$ which encodes only geometric aberrations. Under the eikonal approximation, the transmission function is simply the Fourier transform of the aberration function transmitted through an amorphous specimen potential. Then the label of emittance can be calculated directly from Equation 15. Aberration functions for the training data set are randomly generated. The aberrations coefficients are specified in ranges to match those present on a real aberration-corrected STEM. The resultant calculated emittance and defocus are normalized by their maximum value in the training data set in order to span a range of (0,1]. In practice, the training label is set to be the sum of normalized emittance and normalized defocus $\epsilon = (\hat{\varepsilon} + \widehat{C_1})/2$. The reason is that although defocus does not affect emittance, it does affect the appearance of the Ronchigram. Correcting defocus at the same time helps with the validation on real microscopes. Ronchigrams are simulated with 1,024 × 1,024 px size without an aperture, then as a postprocessing step cropped to 128 × 128 px to match the size of an EMPAD [42] and normalized to the maximum pixel value. The aperture is applied later in training with the size according to the real microscope case by case. Details of the training data and training parameters used can be referred to in Table I and II in Appendix A.

## III. RESULTS

### A. CNN robustness to acquisition conditions

We validate the CNN performance against three common varying factors in the acquisition that may affect the appearance of the Ronchigrams: (i) Shift of the center of the beam; (ii) Range of collection angle; (iii) Exposure time. Results are presented in Figure 3. Overall, the CNN is robust against the three factors. Unsurprisingly, shift of the center of the beam barely affects the CNN performance as the CNN preserves local translational invariance. The CNN overestimates emittance when the collection angle is smaller and underestimates emittance when the collection angle is larger. This is easily explainable as zooming in is equivalent to having a larger flat center area in the field of view which translates to higher emittance and vice versa. The CNN performance degrades as the exposure time becomes shorter, which is also expected as the higher Poisson noise damages the recognition of the flat center area of the Ronchigram.

Nevertheless, the CNN response is able to remain monotonic as a function of ground truth. As long as the monotonicity holds, the Ronchigram – emittance growth pairs can be optimized to reach optimum.

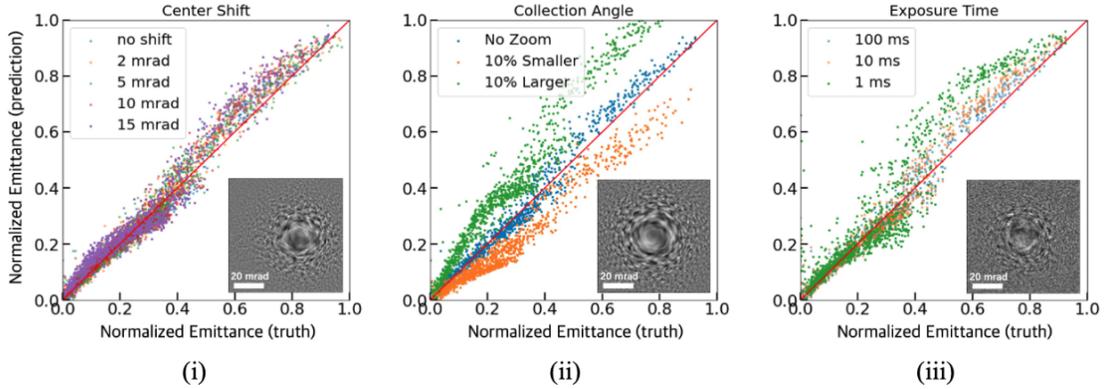

(i)                        (ii)                     (iii)

FIG. 3: Validation of CNN robustness against changes in acquisition conditions: (i) Shift of the center of the beam; (ii) Range of collection angle; (iii) Exposure time. The CNN trained with 25,000 Ronchigrams easily can tolerate lateral shifts up to 15 mrad, 10% difference in collection angle and 1% of the ordinary exposure time.

### B. Beam emittance measurement on real microscopes

We validated our derived emittance metric and customized CNN on real Ronchigrams collected from two aberration-corrected STEMs: ThermoFisher Titan Cryo-S/TEM and ThermoFisher Spectra 300. For both microscopes, control of the camera is made possible by the AutoScript TEM Software[1]. We set the accelerating voltage to be 300 kV for both and semi-convergence angle to be 21.4 mrad and 30.0 mrad, respectively. The aberration correctors are controlled via an RPC protocol provided by CEOS. We collected Ronchigrams with first-order aberrations ranging from -200 nm to 200 nm and second order aberrations from -5 $\mu$m to 5 $\mu$m, as set by the calibrated "knobs" provided by the aberration corrector software. Within each acquisition, the pixel number is set to be 2048 by 2048 and the acquisition time is set to be 800 ms. We apply the customized CNN to predict their corresponding beam emittance growth values and plot against the aberration coefficients. Figure 4 shows the CNN response to varying one aberration coefficient at each direction at a time. Figure 5 shows the CNN response to varying one aberration coefficient at both x and y directions. The valley-shaped curves indicate that our CNN accurately captures the increase of emittance as a result of increased aberration, the symmetry about the origin, as well as the convexity near the optimum. We discuss the general optimization of multiple microscope parameters simultaneously using Bayesian optimization in part II of the paper.

---

[1] AutoScript: https://www.thermofisher.com/us/en/home/electron-microscopy/products/software-em-3dvis/autoscript-tem-software.html

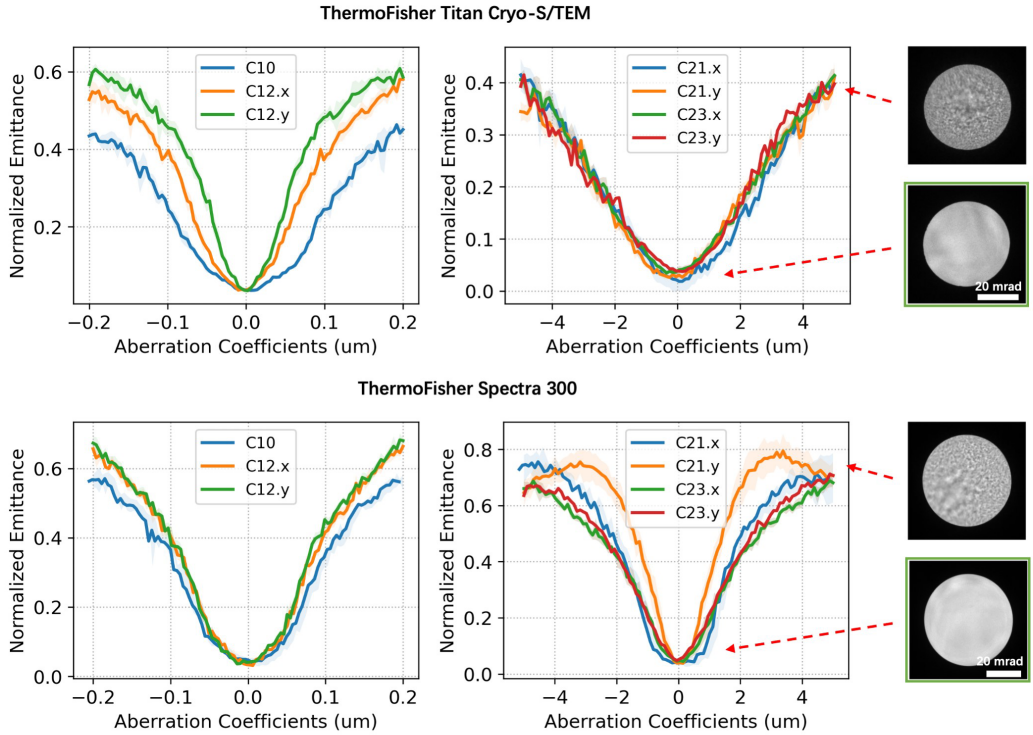

FIG. 4: CNN response to change in single aberration coefficients from ThermoFisher Titan Cryo S/TEM (Upper) and ThermoFisher Spectra 300 (Lower) line scans (acquisition time 800 ms). Left: First order aberrations. Right: Second order aberrations. The CNN accurately captures the increase of emittance as a result of increased aberration. Note how the C21.y of Spectra 300 (orange line) looks different from the others, as it is a known issue with our installation on this particular column. It narrows the range for tuning but still guarantees the optimum will be found.

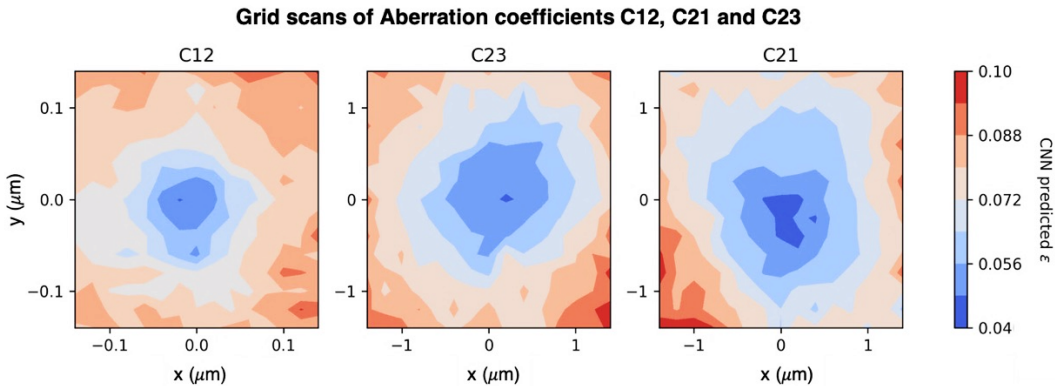

FIG. 5: CNN response to change in single aberration coefficients from 15 by 15 grid scans on ThermoFisher Spectra 300 (acquisition time 800 ms). The convexity at center verifies the existence of optimum.

Again we stress that though defocus $C_{1,0}$ should not affect emittance itself, our CNN is trained on the sum of beam emittance growth and defocus, which we call the normalized emittance, as it does affect the appearance of the Ronchigram. The normalized emittance is not exactly zero at origin, because as discussed earlier the beam can suffer from intrinsic instability and higher order aberrations are inevitable. Nevertheless, we can summarize the CNN is able to accurately read the emittance growth induced by aberrations from the Ronchigrams and can serve as a reliable single-value metric to characterize the probe quality of aberration-corrected STEMs and be used as the objective function in the optimization of aberration correctors.

## IV. DISCUSSION

This work provides a new approach to beam quality evaluation and aberration correction in microscopy. We first introduce *beam emittance* from the field of accelerator physics to the electron microscopy community, by deriving the fundamental connection between the aberration function of a microscope and the induced beam emittance growth in phase space via Wigner distribution of the probe state. Thus, aberration correction is equivalent to emittance minimization. The lack of connection between the two fields is surprising considering the electron microscope is essentially a room sized particle accelerator. The derived beam emittance growth holds several nice properties, (i) Single-valued. (ii) Bounded between (0,1). (iii) Convexity of aberration coefficients. (iv) Independent from defocus. These properties make it ideal for a new beam quality metric in tuning the microscope and more robust objective for optimization.

Ideally, beam emittance minimization will simultaneously minimize all aberration coefficients. However, in practice it is common that certain higher order aberrations are constrained. Another advantage of the emittance approach is that it naturally leads to the compensation of higher order aberrations with lower orders, similar to the Scherzer condition for balancing spherical aberration with defocus. For instance, in some earlier corrector designs $C_{5,0}$ could not be reduced below ~5mm [21]. For the example of $C_{5,0}$ as a limiting aberration, the minimum value of emittance growth occurs when an offsetting spherical aberration of $\tilde{C}_{30}^* = \frac{-9\alpha_0^2}{15}\tilde{C}_{50}$ is present, where $\alpha_0$ is the aperture size. It is also possible to derive the optimal aperture if we include the source emittance $\varepsilon_i$ into the equation and perform brightness maximization. For example, in the case of nonzero $\tilde{C}_{30}$, we reach $\alpha_{0|\tilde{c}_{30}} = \sqrt[4]{\frac{3\sqrt{2}\varepsilon_i}{\pi} \frac{\lambda}{|\tilde{C}_{30}|}}$ where we may interpret $\varepsilon_i$ as the phase tolerance introduced by the source. The derivation of these conditions is given in Appendix C. It is worth noting that, since emittance growth is independent of defocus, the optimal conditions derived are usually different from the ordinary Scherzer conditions by a prefactor [43].

In order to apply this metric to the problem of optimization of real microscopes, we developed a supervised machine learning approach to extract beam emittance growth experimentally from electron Ronchigrams. Our convolutional neural network (CNN) is trained with ground-truth emittance growth values so that it can predict the emittance growth from a single new Ronchigram. This new probe diagnosis method bypasses the measurement of individual aberration coefficients, and thus executes in a few milliseconds once the Ronchigram is acquired whereas the conventional Zemlin tableau takes approximately 2 minutes on average. This is also in contrast to existing machine learning approaches to aberration correction that aim at estimating the aberration coefficients first. We validate the robustness of the trained CNN against errors in acquisition conditions and show the predictions remain reliable within 50% of center shift, 10% of collection angle mismatch and up to 1% of ordinary exposure time.

This shows that machine learning models with sufficiently large amount of training data can accurately capture the correlations between quantities in a complex physical system. In fact, our work might present as a test field for interpretable AI for science and potentially discovering new physics. For example, the currently commonly used aberration function in Equation 2 is expanded under the isoplanatic approximation. The true expression is mathematical challenging and awaits exploration but might be revealed by machine learning with big data. Nonetheless, a key limitation of supervised learning is the construction of a comprehensive training data set. This can lead to scalability concerns when applying the model to different instruments. A promising future direction could be a combination with unsupervised learning and few-shot learning.

Our results show this machine learning model to predict emittance growth has the necessary properties and performance to serve as a reliable and fast-executing function to realize automated aberration correction for electron microscopes. This is a significant departure from conventional aberration analysis and correction protocols, which aim to estimate the aberration function and thus requires a large amount of time and computational resources in the acquisition of real space images of calibration specimen or Ronchigrams, and still require qualitative human adjustment along the way of the tuning. This paper sheds light on full automation of aberration corrector tuning on STEMs and can greatly free the process from heavy labor cost and human bias. Meanwhile, this work resonates with recent progress in the automation of other complex scientific instruments such as the free-electron laser [44] and synchrotron light sources [45], which calls for more future collaborations. Future directions of work would be better optimizers and the possibility of leveraging the physics in the optimizer itself.

## V. CONCLUSION

The resolution of electron microscopes has been revolutionized by the development of aberration correction hardware. However, the intricate design of electromagnetic lenses in these correctors places a substantial demand on precise and rapid software tuning. Current techniques are limited, as they rely on estimating the aberration function—a process that is both time-consuming and computationally intensive, requiring the correlation of extensive sets of images in either real or diffraction space. Additionally, these methods do not directly yield the necessary electric currents to optimize lens performance, and as a result the alignment and calibration of the tuning software can be a time-consuming process during installation, and requires periodic retuning and calibration by the microscope company to keep the instrument in an optimal state.

In this work, we introduced a novel approach to microscope optimization that combines principles from accelerator physics with machine learning techniques. We proposed alternative metrics that surpass the traditional aberration function for assessing electron probe quality. Specifically, we (i) theoretically derived the beam emittance growth caused by aberrations, establishing it as a unified and stable metric for evaluating beam quality, (ii) train a deep neural network to enable near-instant assessment of beam quality by predicting emittance growth from electron Ronchigrams, and (iii) demonstrated that the trained model can be used as a rapid and efficient function for global optimization of lens parameters.

Part I of this paper lays the groundwork for a fast, accurate, and more automated alignment processes for electron microscopes by providing a stable metric. In Part II, we further explore how to achieve online optimization for real-world microscopes involving multiple parameters simultaneously by employing Bayesian optimization. This approach is benchmarked against conventional aberration correction methods, illustrating its potential to enhance the speed and precision of microscope tuning.

**ACKNOWLEDGMENTS**

We thank John Grazul and Mariena Sylvestry Ramos for technical support and maintenance of the electron microscopes. We thank Heiko Müller and his colleagues at CEOS for helping us access their corrector optics. This work was funded by the Center for Bright Beams, an NSF STC (NSF PHY-1549132). Electron Microscopy facilities support from the Cornell Center for Materials Science, an NSF MRSEC supported by DMR-1719875, and the PARADIM Materials Innnovation Platform (NSF DMR-2039380)

**Appendix A: Training the CNN**

We apply the open source Ronchigram simulator[2] [40] to generate random Ronchigrams to be used as our training data set. In order to make accurate one-to-one predictions of emittance, we aim to simulate all possible appearances of Ronchigrams by setting wide ranges of aberration coefficients, while keeping the distribution mostly uniform instead of skewed. A complete list of all aberration coefficient limits used in the simulation are shown in Table II. The total data set contains 25,000 Ronchigrams and split by a 0.8 ratio for
training and testing respectively.

The Ronchigrams are simulated with 1,024 × 1,024 px size without an aperture in a reciprocal space limit of 50 mrad, then as a postprocessing step cropped to 128 × 128 px to match the size of an EMPAD and normalized to the maximum pixel value. No aperture is
applied until the training stage. We leave two options of training: (i) Single-stage training: Apply an aperture corresponding to the microscope before training. (ii) Two-stage training: First do coarse-training without an aperture applied for the deep learning model to capture all phase space variations within the simulation limit, then do fine-training with the aperture applied. In this paper we have adopted the latter approach. Random rotation by 90 degrees is applied in both cases.

The hyperparameter setting for the VGG-16 model is displayed in Table III.

TABLE I: Aberration limits in the simulated Ronchigrams data set.

| Aberration Coefficient | Symbol[a] | Unit | 3-fold | C5+(-C1) | C3+(-C1) | Random |
|---|---|---|---|---|---|---|

---

[2] http://ronchigram.com/

| Aberration | Symbol | Unit | | | | |
|---|---|---|---|---|---|---|
| Defocus | $C_{10}$ | Å | [-1000,1000] | [-1200,0] | [-1200,500] | [-10000,10000] |
| 2-Fold Astigmatism | $C_{12}$ | nm | 0 | 0 | 0 | [-1250,1250] |
| Axial Coma | $C_{21}$ | nm | 0 | 0 | 0 | [-2500,2500] |
| 3-Fold Astigmatism | $C_{23}$ | nm | [-6000,6000] | [0,1000] | [0,1000] | [-2500,2500] |
| 3rd Order Spherical | $C_{30}$ | μm | [-900,900] | [-100,0] | [200,1000] | [-1000,1000] |
| 3rd Order Axial Star | $C_{32}$ | μm | 0 | 0 | 0 | [-200,200] |
| 4-Fold Astigmatism | $C_{34}$ | μm | 0 | 0 | 0 | [-50,50] |
| 4th Order Axial Coma | $C_{41}$ | mm | 0 | 0 | 0 | [-0.75,0.75] |
| 3-Lobe Aberration | $C_{43}$ | mm | 0 | 0 | 0 | [-0.05,0.05] |
| 5-fold Astigmatism | $C_{45}$ | mm | 0 | 0 | 0 | [-0.25,0.25] |
| 5th Order Spherical | $C_{50}$ | mm | [-40,40] | [20,400] | 0 | [-200,200] |
| 5th Order Axial Star | $C_{52}$ | mm | 0 | 0 | 0 | 0 |
| 5th Order Rosette | $C_{54}$ | mm | 0 | 0 | 0 | 0 |
| 6-fold Astigmatism | $C_{56}$ | mm | 0 | 0 | 0 | 0 |
| Chromatic Aberration | $C_c$ | m | 0 | 0 | 0 | 0 |

[a] Krivanek Notation

TABLE II: Hyperparameter setting for the VGG-16 model

| Hyperparameter | Value |
|---|---|
| optimizer | Adam |
| learning rate | [1e-4, 5.5e-6] |
| betas | [(0.9, 0.999), (0.9, 0.999)] |
| eps | [1e-7, 1e-7] |
| epoch | [100, 100] |
| batch size | 32 |
| patience | 5 |
| min delta | 1e-4 |
| split | 0.8 |
| dropout | 0.3 |

**Appendix B: CNN validation on the Nion UltraSTEM**

At an earlier stage of the project, we also tested on a Nion UltraSTEM microscope before it was decommissioned. The Nion microscope used a larger aperture size of 45 mrad which would include more higher order aberrations in the Ronchigram, but also had some intrinsic limitations in the lens design that prohibited perfect tuning even at smaller angles. Nevertheless, the CNN performs equally well on this microscope, showing clear convexity near the optimum.

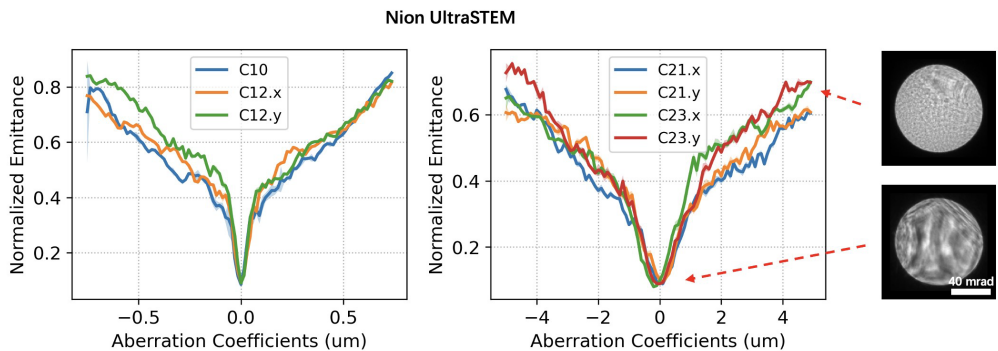

FIG. B1: CNN response to change in single aberration coefficients from Nion UltraSTEM line scans (acquisition time 200 ms). Upper: First order aberrations. Lower: Second order aberrations. The CNN accurately captures the increase of emittance as a result of increased aberration.

**Appendix C: From emittance minimization to Scherzer conditions**

Although ideally beam emittance minimization should simultaneously minimize all the aberration coefficients, we show it can still reach an optimal state even when there are constraints on higher order aberrations by compensating them with the lower orders similar to the Scherzer approach.

Inspect the form of emittance growth in Equation 12,

$$\varepsilon_\chi^2 = \frac{4\pi^2}{\lambda^2} \sum_{nn'm} \Re[\tilde{C}_{n,m}^* \tilde{C}_{n',m}] \alpha_0^{n+n'+2} \left( \frac{1}{n+n'+2} + \frac{m^2}{(n+1)(n'+1)(n+n'+2)} - \frac{4\delta_{m0}}{(n+3)(n'+3)} \right) \quad (12)$$

Consider the case of compensating uncorrectable $\tilde{C}_{5,0}$ with $\tilde{C}_{3,0}$.

$$\varepsilon_{\tilde{C}_{30},\tilde{C}_{50}}^2 = \frac{4\pi^2}{\lambda^2} \left( \frac{|\tilde{C}_{30}|^2 \alpha_0^8}{72} + \frac{\Re[\tilde{C}_{30}^* \tilde{C}_{50}] \alpha_0^{10}}{60} + \frac{|\tilde{C}_{50}|^2 \alpha_0^{12}}{48} \right) \quad (C1)$$

The emittance growth is minimized at

$$\frac{d\, \varepsilon_{\tilde{C}_{30},\tilde{C}_{50}}^2}{d\tilde{C}_{30}} = 0 = \frac{4\pi^2}{\lambda^2} \left( \frac{\tilde{C}_{30}^* \alpha_0^8}{72} + \frac{\tilde{C}_{50} \alpha_0^{10}}{120} \right) \quad (C3)$$

which yields

$$\tilde{C}_{30}^* = \frac{-9\alpha_0^2}{15} \tilde{C}_{50} \quad (C4)$$

It is also possible to derive the optimal aperture. Consider nonzero $\tilde{C}_{30}^*$,

$$\varepsilon_{\tilde{C}_{30}}^2 = \frac{4\pi^2}{\lambda^2} \frac{|\tilde{C}_{30}|^2 \alpha_0^8}{72} \quad (C5)$$

The brightness of the beam including the source emittance $\varepsilon_i^2$ is

$$B_{\tilde{C}_{30}} = \frac{\pi J \alpha_0^2}{\sqrt{\frac{4\pi^2}{\lambda^2} \frac{|\tilde{C}_{30}|^2 \alpha_0^8}{72} + \varepsilon_i^2}} \quad (C6)$$

Optimal aperture is reached by brightness maximization,

$$\frac{dB_{\tilde{C}_{30}}}{d\alpha_0} = 0 = \frac{\pi 2 J \alpha_0 \sqrt{\frac{4\pi^2}{\lambda^2} \frac{|\tilde{C}_{30}|^2 \alpha_0^8}{72} + \varepsilon_i^2} - \frac{1}{2} \frac{\pi J \alpha_0^2 \frac{4\pi^2}{\lambda^2} \frac{|\tilde{C}_{30}|^2 8\alpha_0^7}{72}}{\sqrt{\frac{4\pi^2}{\lambda^2} \frac{|\tilde{C}_{30}|^2 \alpha_0^8}{72} + \varepsilon_i^2}}}{\frac{4\pi^2}{\lambda^2} \frac{|\tilde{C}_{30}|^2 \alpha_0^8}{72} + \varepsilon_i^2} \quad (C7)$$

which yields

$$\alpha_{0|\tilde{C}_{30}} = \sqrt[4]{\frac{3\sqrt{2}\varepsilon_i}{\pi}\frac{\lambda}{|\tilde{C}_{30}|}} \tag{C8}$$

Similarly, if we consider nonzero $\tilde{C}_{50}^*$, brightness maximization yields,

$$\alpha_{0|\tilde{C}_{50}} = \sqrt[6]{\frac{6\varepsilon_i}{\pi}\frac{\lambda}{|\tilde{C}_{50}|}} \tag{C9}$$

Therefore, beam emittance minimization approach also provides an alternative way to select the optimal aperture besides the Scherzer conditions, if we understand the intrinsic non-dimensional emittance of our source $\varepsilon_i$ as related to the phase tolerance in Scherzer's approach, $\varphi$ (traditionally set at $\pi/4$) [43,46].

However, it is analytical infeasible to solve for optimal aperture under multiple higher order aberrations. For example, consider brightness maximization under nonzero $\tilde{C}_{30}^*$ and $\tilde{C}_{50}^*$,

$$\varepsilon_{\tilde{C}_{30},\tilde{C}_{50}}^2 = \frac{4\pi^2}{\lambda^2}\left(\frac{|\tilde{C}_{30}|^2 \alpha_0^8}{72} + \frac{\Re[\tilde{C}_{30}^*\tilde{C}_{50}]\alpha_0^{10}}{60} + \frac{|\tilde{C}_{50}|^2 \alpha_0^{12}}{48}\right) \tag{C10}$$

which solves high degree polynomials beyond 6,

$$\frac{|\tilde{C}_{30}|^2 \alpha_0^8}{18} + \frac{\Re[\tilde{C}_{30}^*\tilde{C}_{50}]\alpha_0^{10}}{10} + \frac{|\tilde{C}_{50}|^2 \alpha_0^{12}}{6} = \frac{\varepsilon_i^2 \lambda^2}{\pi^2} \tag{C11}$$

We can verify that $\alpha_{max}$ will have the correct dependence on $(\tilde{C}_{30})^{-\frac{1}{4}}$ and $(\tilde{C}_{50})^{-\frac{1}{6}}$ and numerically solve for $\alpha_{0|\tilde{C}_{30},\tilde{C}_{50}}$.

**Appendix D: A closer look at the quantum definition of beam emittance**

The classical RMS definition of emittance is given in Equation 1

$$\varepsilon^2 = \langle x^2\rangle\langle p^2\rangle - \langle x \cdot p\rangle^2 \tag{D1}$$

where position and momentum of the electron are assumed to be measured simultaneously.

In the quantum regime, we should similarly define emittance from expectations of quantum operators, but symmetrize the cross term because the $x$ and $p$ as operators do not commute. A proper way of doing it is by defining an observable $\hat{A} = (\hat{x}\hat{p} + \hat{p}\hat{x})/2$ so that,

$$\varepsilon^2 = \langle \hat{x}^2\rangle\langle \hat{p}^2\rangle - \langle (\hat{x}\hat{p} + \hat{p}\hat{x})/2\rangle^2 \tag{D2}$$

Now we examine the time evolution of $\varepsilon^2$. In free space, we should have $\frac{d\varepsilon^2}{dt} = 0$.

The Ehrenfest theorem relates the time derivative of the expectation of any quantum mechanical operator to the expectation of the commutator of that operator with the Hamiltonian of the system. Therefore we can calculate the time derivate of each term on the right hand side of Eqaution C2 by,

$$\frac{d\langle \hat{A}\rangle}{dt} = \frac{i}{\hbar}\langle[\hat{H},\hat{A}]\rangle + \left\langle\frac{\partial A}{\partial t}\right\rangle, \qquad H = \frac{\hat{p}^2}{2m} \tag{D3}$$

where in our case $\left\langle\frac{\partial A}{\partial t}\right\rangle = 0$.

$$\begin{aligned}
\frac{d\langle \hat{x}^2\rangle}{dt} &= \frac{i}{\hbar}\langle[H,\hat{x}^2]\rangle = \frac{i}{2m\hbar}\langle \hat{p}[\hat{p},\hat{x}^2] + [\hat{p},\hat{x}^2]\hat{p}\rangle \\
&= \frac{i}{2m\hbar}\langle \hat{p}(-i\hbar 2\hat{x}) + (-i\hbar 2\hat{x})\hat{p}\rangle \\
&= \frac{1}{m}\langle \hat{x}\hat{p} + \hat{p}\hat{x}\rangle
\end{aligned} \tag{D4}$$

where we use the the canonical relation $[\hat{p},\hat{x}^n] = -i\hbar n \hat{x}^{n-1}$.

$$\frac{d\langle \hat{p}^2\rangle}{dt} = \frac{i}{\hbar}\langle[H,\hat{p}^2]\rangle = \frac{i}{2m\hbar}\langle[\hat{p}^2,\hat{p}^2]\rangle = 0 \tag{D5}$$

Since $\langle \hat{p}^2\rangle$ does not change with time, so we can define $\langle \hat{p}^2\rangle = p_0^2$.

$$\begin{aligned}
\frac{d\langle \hat{A}\rangle}{dt} &= \frac{d\langle \hat{x}\hat{p} + \hat{p}\hat{x}\rangle}{2dt} = \frac{i}{2\hbar}\langle[H,\hat{x}\hat{p} + \hat{p}\hat{x}]\rangle \\
&= \frac{i}{2\hbar}\frac{1}{2m}(\langle[\hat{p}^2,\hat{x}\hat{p}]\rangle + \langle[\hat{p}^2,\hat{p}\hat{x}]\rangle) \\
&= \frac{i}{2\hbar}\frac{1}{2m}(\langle \hat{p}[\hat{p},\hat{x}]\hat{p}\rangle + \langle \hat{p}[\hat{p},\hat{x}]\hat{p}\rangle) \\
&= \frac{i}{2\hbar}\frac{1}{2m}(-2\hbar\,\hat{p}^2) \\
&= \frac{1}{2m}\langle \hat{p}^2\rangle = \frac{p_0^2}{2m}
\end{aligned} \tag{D6}$$

$$\frac{d\langle \hat{A}\rangle^2}{dt} = 2\langle \hat{A}\rangle\frac{d\langle \hat{A}\rangle}{dt} = 2\langle \hat{A}\rangle\frac{p_0^2}{2m} = \frac{p_0^2}{m}\cdot\langle \hat{x}\hat{p} + \hat{p}\hat{x}\rangle \tag{D7}$$

Therefore,

$$\frac{d\varepsilon^2}{dt} = p_0^2\frac{d\langle \hat{x}^2\rangle}{dt} - \frac{d\langle \hat{A}\rangle^2}{dt} = \frac{p_0^2}{m}\langle \hat{x}\hat{p} + \hat{p}\hat{x}\rangle - \frac{p_0^2}{m}\cdot\langle \hat{x}\hat{p} + \hat{p}\hat{x}\rangle = 0 \tag{D8}$$

This shows that the quantum definition of beam emittance is also conserved in free space.